\documentclass[twocolumn,tighten,times]{aastex631}
\received{Nov. 1, 2023}
\revised{Dec. 1, 2023}

\shorttitle{Dwarf pulses of 10 pulsars}
\shortauthors{Y. Yan et al.}
\usepackage{subfigure}

\graphicspath{{Graphs/}}

\usepackage{ulem}
\usepackage{CJKulem}
\usepackage{soul}

\begin{document}

\title{Dwarf pulses of 10 pulsars detected by FAST
}

\correspondingauthor{J.~L. Han} 
\email{hjl@nao.cas.cn (JLH)}


\author[0009-0008-1612-9948]{Yi Yan}
\affiliation{National Astronomical Observatories, Chinese Academy of Sciences,
         Jia-20 Datun Road, ChaoYang District, Beijing 100101, China}
\affiliation{School of Astronomy, University of Chinese Academy of Sciences,
         Beijing 100049, China}

\author[0000-0002-9274-3092]{J.~L. Han}
\affiliation{National Astronomical Observatories, Chinese Academy of Sciences,
         Jia-20 Datun Road, ChaoYang District, Beijing 100101, China}
\affiliation{School of Astronomy, University of Chinese Academy of Sciences,
         Beijing 100049, China}
\affiliation{CAS Key Laboratory of FAST, NAOC, Chinese Academy of Sciences, Beijing 100101, China}

\author[0000-0002-6423-6106]{D.~J. Zhou}
\affiliation{National Astronomical Observatories, Chinese Academy of Sciences,
         Jia-20 Datun Road, ChaoYang District, Beijing 100101, China}
\affiliation{School of Astronomy, University of Chinese Academy of Sciences,
         Beijing 100049, China}

\author[0000-0003-1946-086X]{L. Xie}
\affiliation{National Astronomical Observatories, Chinese Academy of Sciences,
         Jia-20 Datun Road, ChaoYang District, Beijing 100101, China}
\affiliation{School of Astronomy, University of Chinese Academy of Sciences,
         Beijing 100049, China}
         
\author{F.~F. Kou}
\affiliation{Xinjiang Astronomical Observatories, Chinese Academy of Sciences, Urumqi 830011, China}

\author[0000-0002-6437-0487]{P.~F. Wang}
\affiliation{National Astronomical Observatories, Chinese Academy of Sciences,
         Jia-20 Datun Road, ChaoYang District, Beijing 100101, China}
\affiliation{School of Astronomy, University of Chinese Academy of Sciences,
         Beijing 100049, China}
         
\author[0009-0004-3433-2027]{C. Wang}
\affiliation{National Astronomical Observatories, Chinese Academy of Sciences,
         Jia-20 Datun Road, ChaoYang District, Beijing 100101, China}
\affiliation{School of Astronomy, University of Chinese Academy of Sciences,
         Beijing 100049, China}

\author[0000-0002-4704-5340]{T. Wang}
\affiliation{National Astronomical Observatories, Chinese Academy of Sciences,
         Jia-20 Datun Road, ChaoYang District, Beijing 100101, China}
\affiliation{School of Astronomy, University of Chinese Academy of Sciences,
         Beijing 100049, China}

\begin{abstract}
How pulsars radiate is a long-standing problem. Detailed polarization measurements of individual pulses shed light on currently unknown emission processes. Recently, based on supersensitive observations, dwarf pulses have been recognized as weak narrow pulses often appearing during the nulling state. In this study, we report the detection of dwarf pulses from ten pulsars, PSRs B0525+21, B1237+25, J1538+2345, J1824$-$0127, J1851$-$0053, B1901+10, J1939+10, B1944+17, B2000+40 and J2112+4058, based on observations conducted with the Five-hundred-meter Aperture Spherical radio Telescope. Dwarf pulses of five pulsars are clearly discernible in the two-dimensional distribution of pulse intensity and pulse width. For the other five pulsars, PSRs J1538+2345, J1824$-$0127, J1939+10, B2000+40, and J2112+4058, only a few dwarf pulses are detected from pulse stacks. The dwarf pulses can emerge in both cone and core emission components for PSR B1237+25, and the polarization angles of these dwarf pulses are mostly in the orthogonal polarization mode of normal pulses for PSR B1944+17. In general, pulsars with detected dwarf pulses tend to be located within the ``death valley" region of the distribution of pulsar periods and period derivatives.
\end{abstract}


\section{Introduction} \label{sec:intro}

The radio emission of pulsars sometimes shows different modes, in which individual pulses and hence the mean profiles appear different for some periods \citep{Backer1970_ModeChange}. This is called the ``mode changing" of pulsar radio emission.
One mode of some pulsars can be the case that emission is completely quenched, i.e. no detectable emission for some periods, that is ``nulling''  \citep{Backer1970_Nulling}. 
Nulling behavior has been observed in more than 200 pulsars \citep[e.g.,][]{Ritchings1976, Wang2007, Gajjar2012, Basu2017, Wang2020}.
In the conventional model, the radio emission of pulsars is produced by the particles streaming from the polar caps around the magnetic poles of neutron stars \citep{Ruderman1975}. 
Though the details of physical processes for the nulling are not clear, the observational ``nulling'' is found to be related to the particle generation and outflows in the pulsar magnetosphere, since pulsars have different spin-down rates in the on and off states \citep{Kramer2006}.
During the pulse nulling phases, sporadic weak narrow pulses, i.e., the dwarf pulses, have been detected from some pulsars, such as PSRs B1237+25 \citep{Srostlik2005} and J1107$-$5907 \citep{Young2014}.
By integrating nulling pulses, weak emission has been detected from PSRs B0826-34 \citep{Esamdin2005} and B1944+17 \citep{Kloumann2010}.
These imply that the emission may not be completely ceased in the nulling periods.

\begin{deluxetable*}{lrrclccc}
\tablenum{1}
\tablecaption{FAST observations of 10 pulsars with detected dwarf pulses. Column (1-3): pulsar name, period and DM; Column (4-6): FAST  observation date, cover and beam name, and observation time; Column (7-8): number of pulsar periods and the number of dwarf pulses detected.}
\label{table:obs}
\tablehead{
\colhead{PSR Name} & \colhead{Period} & \colhead{DM} & \colhead{ObsDate} &
\colhead{Cover and FAST Beam Name} & \colhead{T$_{\rm obs}$} & \colhead{No. of } & \colhead{No. of } \\
\colhead{} & \colhead{(s)} & \colhead{(cm$^{-3}$pc)} & \colhead{(yyyy/mm/dd)} &
\colhead{} & \colhead{(minutes)} & \colhead{Periods} & \colhead{Dwarf pulses}
}
\decimalcolnumbers
\startdata
B0525+21          & 3.746 & 50.87 & 2022/03/04 & J052856+221135\_M11P1  & 15        & 243         & 12     \\
B1237+25          & 1.382 & 9.25  & 2020/01/18 & J1239+2453\_M01P1*     & 120       & 5210        & 191    \\
J1538+2345 & 3.449 & 14.91 & 2020/12/16 & J1538+2345\_M01P1* & 10 & 171 & 2 \\
                  &       &       & 2021/11/22 & J1538+2345\_M01P1* & 15 & 256  & 4 \\
J1824$-$0127      & 2.499 & 63.0  & 2022/09/02 & J182457-011722\_M11P1  & 15        & 361         & 2      \\
                  &       &       & 2021/08/22 & G28.73+5.08\_M16P2     & 5         & 124         & 1      \\
J1851$-$0053      & 1.409 & 24.0  & 2021/08/22 & G32.20-0.25\_M04P2     & 5         & 220         & 7      \\
B1901+10          & 1.856 & 135.0 & 2023/05/25 & J1904+1011\_M01P1      & 18        & 584         & 11    \\
J1939+10$^{\#}$   & 2.309 & 74.7  & 2021/07/02 & G47.95-5.51\_M08P2     & 5         & 134         & 1      \\
B1944+17          & 0.441 & 16.14 & 2020/10/28 & J1946+1805\_M01P1*     & 55        & 7491        & 234   \\
                  &       &       & 2022/01/02 & J194648+175659\_M17P1  & 15        & 2070        & 84     \\
B2000+40 & 0.905 & 131.49 & 2021/07/17 & G76.72+5.34\_M05P4 & 5 & 367 & 2 \\
J2112+4058  & 4.061 &  129  & 2021/11/29 & G84.79-5.25\_M06P1 & 5 & 76 &  3 \\
\enddata
\tablecomments{\#: New position obtained by multibeam FAST observations is R.A. = 19:39:22.6, decl. = 10:49:54;  *: data from the released FAST archive data.}
\end{deluxetable*}

\citet{Chen2023} recently detected a large number of dwarf pulses in the generally recognized nulling state of PSR B2111+46 using the supersensitive Five-Hundred-Meter Aperture Spherical radio Telescope  \citep[FAST;][]{Nan+2006ScChG..49..129N}. They found that these dwarf pulses are a distinct population from the normal pulses in the two-dimensional distribution of pulse width and pulse intensity. They suggested that these dwarf pulses are likely produced by one or few ``raindrops" of particles generated in a fragile gap, while the normal pulses are produced by copious particles in the ``thunderstorm''.

We are currently using the FAST to carry out the Galactic Plane Pulsar Snapshot (GPPS) survey \citep{Han2021}. During survey observations or the follow-up verification observations for pulsar candidates, known pulsars are often detected by the FAST. Based on mostly the GPPS survey data but also some released FAST archive data, we check individual pulses of known pulsars, and serendipitously find dwarf pulses in the nulling state of 10 other pulsars, in addition to PSR B2111+46. Here we present the properties and polarization of dwarf pulses detected in the nulling state for  these pulsars. Observation and data processing are briefly presented in Section~\ref{sec:Obs}. In Section~\ref{sect:Resu}, we report the results for dwarf pulses in detail. Discussion and conclusions are presented in Section ~\ref{sect:Discu}.

\section{Observations and data reduction} \label{sec:Obs}

Parameters of FAST observed pulsars and relevant details of  observations are listed in Table~\ref{table:obs}. The data of most pulsars are taken from the GPPS survey \citep{Han2021}. Data of PSR B1901+10 are taken from a FAST project to study the narrow pulses of nulling pulsars proposed by the first author of this paper. Data of PSRs J1538+2345, B1237+25 on 2020 January 18 and B1944+17 on 2020 October 28 are taken from the FAST released archive, labeled by * in the column (5) of  Table~\ref{table:obs}. The FAST observations were carried out by using the L-band receiver, covering the frequency range from 1000 to 1500 MHz with 2048 or 4096 channels. The sampling time is 49.152 $\mu$s and full polarization signals were recorded.

We fold the data to obtain both integrated profiles and single-pulse stacks using the DSPSR package \citep{van2011}. The PSRCHIVE software \citep{Hotan2004} is used to estimate and correct the Faraday rotation measure. Additionally, the radio frequency interference has also been excised. 

\begin{figure*}
\centering
\begin{minipage}{0.3\linewidth}
    \centering
    \includegraphics[width=0.99\linewidth]{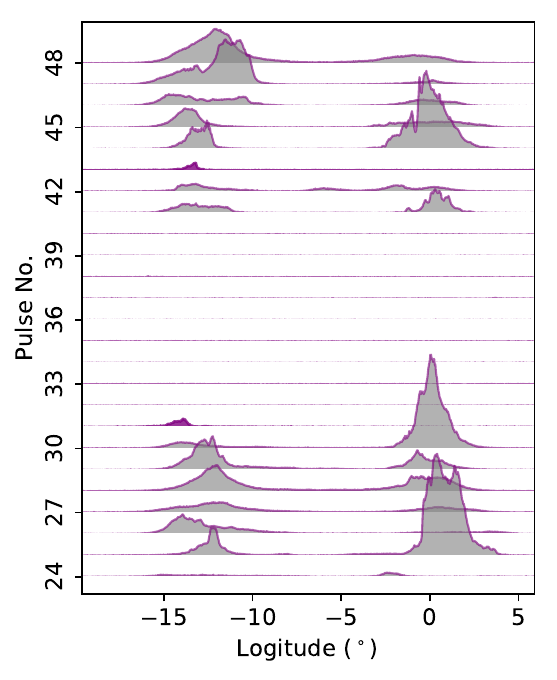}
\end{minipage}
\begin{minipage}{0.68\linewidth}
    \centering
    \includegraphics[width=0.99\linewidth]{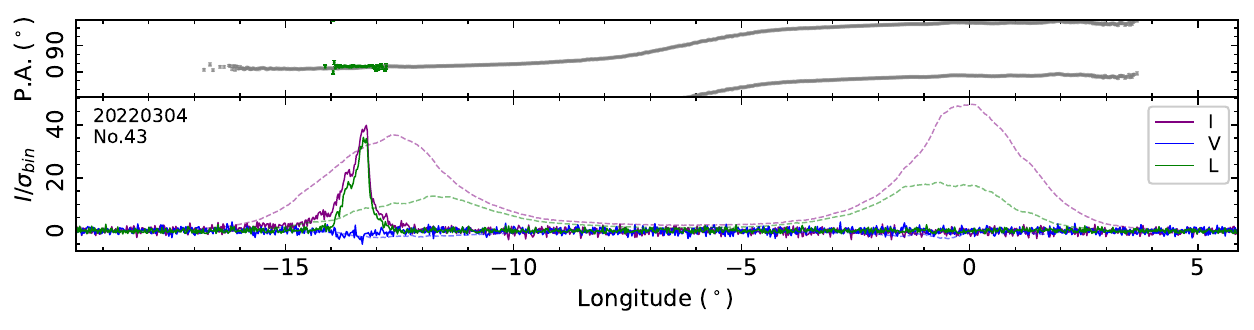}
    \includegraphics[width=0.99\linewidth]{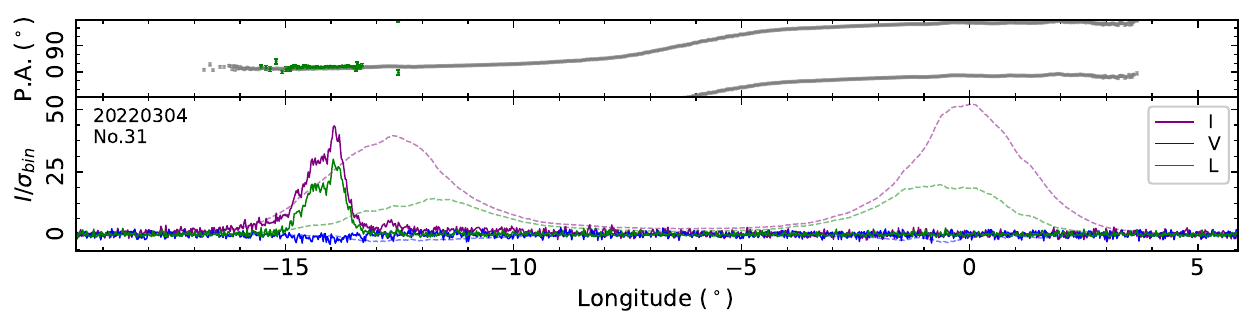}
\end{minipage}
\caption{Dwarf pulses and polarization profiles of PSR B0525+21 observed by FAST on 2022 March 4. A part of single pulse stack, pulses Nos. 24-48, is shown in {\it the left panel}. {\it The right panels} illustrate polarization profiles for the averaged pulses (in dashed fainter lines) and single pulses (solid lines)  with a time resolution of 0.23 ms (16,384 phase bins per period). The linear polarization position angle (PA) curves and data points are plotted in {\it the upper subpanel}, and the total intensity $I$, linear polarization $L$ and circular polarization $V$ are {\it in the lower subpanel}, and they are scaled with $\sigma_{\rm bin}$ that is obtained from the off-pulse fluctuations.
The PA track is plotted twice, once with the original PA value and once with the ``PA value + 180".
}
\label{FigSinPulStack_J0528_20220304}
\end{figure*}

\begin{figure}
\centering
\setlength\tabcolsep{0pt}
\begin{tabular}{ccc}
\includegraphics[width=0.4\textwidth]{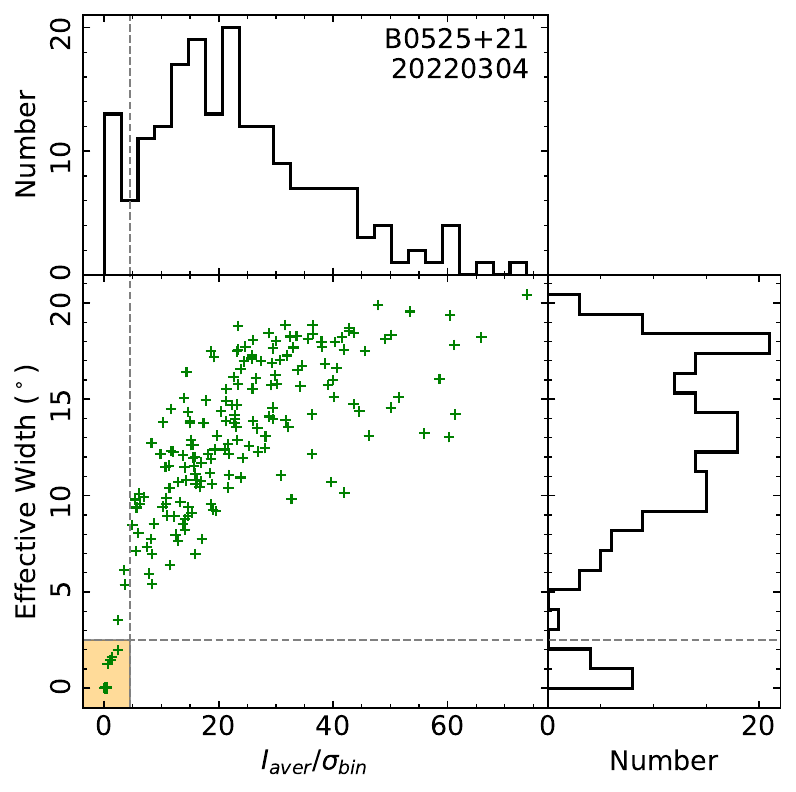}
\end{tabular}
\caption{Distribution of effective pulse width and pulse intensity for single pulses of PSR B0525+21 observed by FAST on 2022 March 4. Similar to PSR B2111+46 by \citet{Chen2023}, the dwarf pulses are distinctly distributed from normal pulses, and have an independent peak on distributions of pulse width and pulse intensity.} 
\label{FigIW_J0528_20220304}
\end{figure}

\begin{figure}
\centering
\setlength\tabcolsep{0pt}
\begin{tabular}{ccc}
\includegraphics[width=0.4\textwidth]{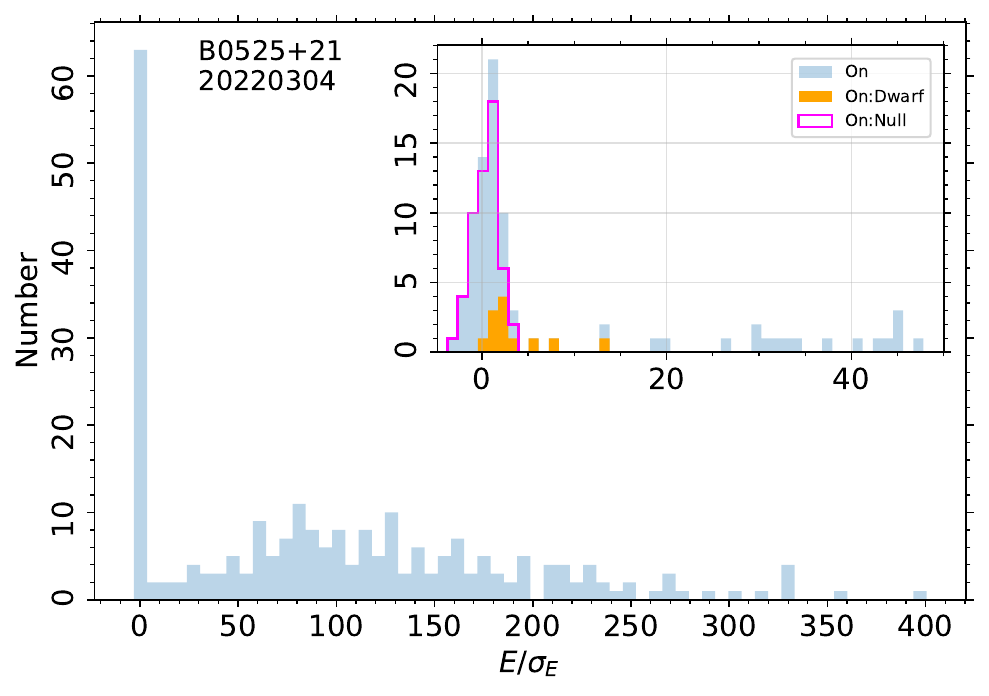}
\end{tabular}
\caption{Energy histogramaaaa of all individual pulses of PSR B0525+21 observed by FAST on 2022 March 4. As shown in the insert, the dwarf pulses have a small energy buried in the histograms for nulling periods, not in the tail for normal pulses.}
\label{FigIhist_J0528_20220304}
\end{figure}

\begin{figure}
\centering
\includegraphics[width=0.42\textwidth]{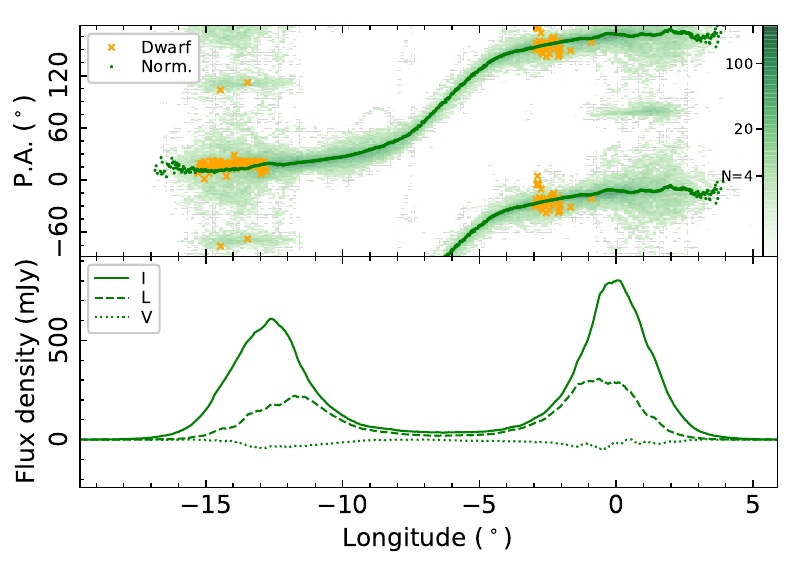}
\caption{
The mean profiles of total intensity $I$ (solid line), linear polarization $L$ (dashed line) and circular polarization $V$ (dotted line) of normal pulses for PSR B0525+21 observed by FAST on 2022 March 4, with the PA distributions. As two dwarf pulses in Figure~\ref{FigSinPulStack_J0528_20220304}, the PA distribution of all dwarf pulses are marked by orange `x' if linear polarization intensity greater than 5$\sigma_{\rm bin}$, which indicates the longitude ranges of dwarf pulse detection. } 
\label{FigPolCon_J0528_20220304}
\end{figure}


The integrated profiles are taken as a pulse emission window. The boundaries of detected single pulses are then determined based on the real signal detection in three successive phase bins exceeding 3$\sigma_{\rm bin}$, or two bins exceeding 5$\sigma_{\rm bin}$, or one bin exceeding 8$\sigma_{\rm bin}$. Here, $\sigma_{\rm bin}$ represents the standard deviation of the off-pulse intensity of single-pulse data. Some very narrow subpulses have one or two phase bins only. The outermost leading and trailing phases are taken to be the phase boundaries of single pulses. When no emission is detected by FAST for these bright pulsars, the pulsar is conventionally believed to be in the null state.

\section{Results and analysis}
\label{sect:Resu}

We then identify dwarf pulses as being sporadic, weak and narrow pulses, as done by \citet{Chen2023}. Here, the effective emission width of a given single pulse is defined by the number of phase bins exceeding 5$\sigma_{\rm bin}$. In general, dwarf pulses are well separated from normal pulses in the distribution of pulse width and pulse intensity, as a distinct single-pulse population \citep{Chen2023}.  
Following the work on PSR J2113+4644 by \citet{Chen2023}, we detect dwarf pulses using FAST from other 10 pulsars, as listed in Table~\ref{table:obs}, and present the emission and polarization properties of dwarf pulses in detail in the following subsections. We present results for five pulsars (PSRs B0525+21, B1237+25, J1851$-$0053, B1901+10 and B1944+17) with many dwarf pulses detected, and for the other five pulsars (PSRs J1538+2345, J1824$-$0127, J1939+10, B2000+40 and J2112+4058) with few dwarf pulses by FAST observations.

\subsection{PSR B0525+21}

PSR B0525+21 is a very bright pulsar discovered by \citet{Staelin1968}.
The pulsar is well known to have nulling \citep{Ritchings1976, Herfindal2009,Basu2017,Wang2020}, mode changing \citep{Basu2021} and subpulse drifting \citep{Weltevrede2006,Weltevrede2007} behaviors. 
\citet{Backer1973} found a long-period modulation about 40 rotation periods for all longitude bins across the pulse. \citet{Herfindal2009} found two harmonically related modulation features related to nulls, and suggested that the quasiperiodic nulling may be produced by the empty passes of the sight line through the carousel beam pattern. This pulsar has two main profile components connected with a bridge emission. The emission geometry has been derived by \citet{Timothy2019}, which has a magnetic incline angle $\alpha$ of 21$^\circ$ and the sight-line impact angle $\beta$ of +0.$^\circ$6. The two main profile components are conal emission \citep{Young2012}.

We obtain data of PSR B0525+21 from the verification observation of a pulsar candidate in the FAST GPPS survey. The observation was conducted for 15 minutes on 2022 March 4, and 12 dwarf pulses are detected. Figure~\ref{FigSinPulStack_J0528_20220304} shows a part of single-pulse stack from pulses Nos. 24-48, together with polarization profiles for the averaged pulse and two dwarf pulses, pulse Nos. 31 and 43. 

Though the data are not so rich due to the short observation, the distributions of the averaged intensity and the effective width of single pulses in Figure~\ref{FigIW_J0528_20220304} still indicate the probably distinct population of dwarf pulses with an averaged intensity of less than 4.5$\sigma_{\rm bin}$ and an effective width of less than 2$^\circ$.5. The energy distribution for all single pulses is depicted in  Figure~\ref{FigIhist_J0528_20220304}, showing an obvious bimodal shape and the other peak for dwarf pulses. The energy is the integrated area under the single-pulse profile and is scaled by $\sigma_{\rm E}$, which represents the standard deviation of the off-pulse energy obtained in the same size of off-pulse phase window. The lower end of this distribution is dominated by dwarf pulses and nulls. 
%

A dwarf pulse may have structured subpulses, as shown by the pulse No. 31 in Figure~\ref{FigSinPulStack_J0528_20220304}. The longitude distribution for 12 dwarf pulses is shown in the PA subpanel of Figure~\ref{FigPolCon_J0528_20220304}, together with the polarization properties for normal pulses. In the case of PSR B0525+21, dwarf subpulses are concentrated in the leading part of two main profile components. The polarization angles of dwarf pulses are consistent with the PA distribution of normal pulses. Although the emission is dominated by the primary polarization mode across the profile for both normal and dwarf pulses, a small number of normal pulses have polarization angles in the orthogonal polarization mode in the outer half longitude range of the two profile components from the conal emission \citep{Timothy2019}. Similar to the results in \citet{Young2012}, we do detect weak subpulses in the bridge of the two components (see pulse No. 42 in Figure~\ref{FigSinPulStack_J0528_20220304}), which are supposed to come from a core component by \citet{Young2012}, while these subpulses are not dwarf pulses according to their pulse width and intensities.

\begin{figure*}
    \subfigure{
        \begin{minipage}{0.3\linewidth}
            \centering
            \includegraphics[width=0.99\linewidth]{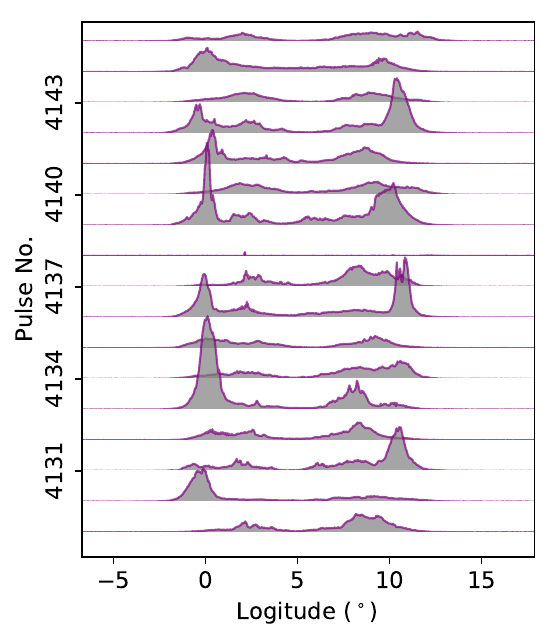}
        \end{minipage}
        \begin{minipage}{0.68\linewidth}
            \centering
            \includegraphics[width=0.95\linewidth]{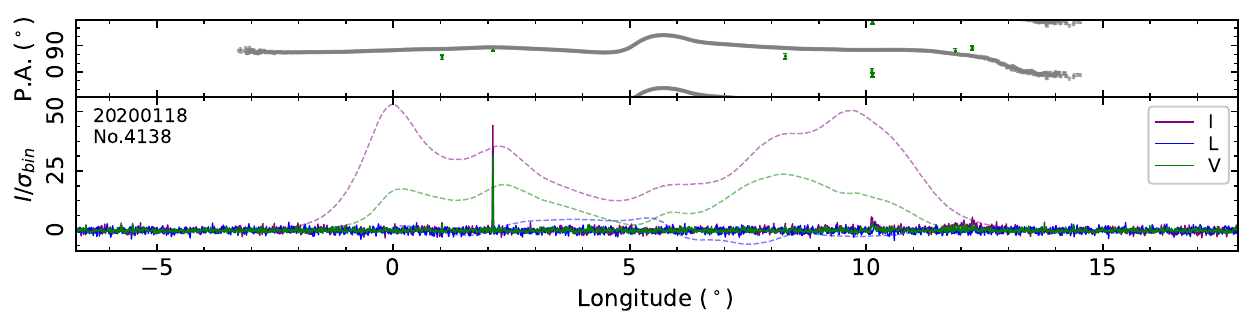}
            \includegraphics[width=0.95\linewidth]{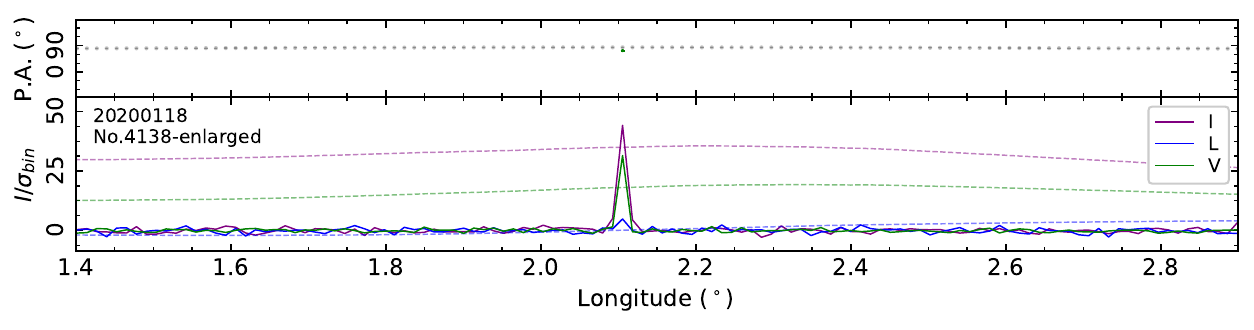}
        \end{minipage}
        }
    \caption{The same as Figure~\ref{FigSinPulStack_J0528_20220304}, but for a single dwarf pulse for PSR B1237+25 observed by FAST on 2020 January 18. Polarization profiles have a time resolution of 49.152 $\mu$s, with an enlarged version in the panel below. }
    \label{FigSinPulStack_J1239}
\end{figure*}

\begin{figure}
\centering
\setlength\tabcolsep{0pt}
\begin{tabular}{ccc}
\includegraphics[width=0.4\textwidth]{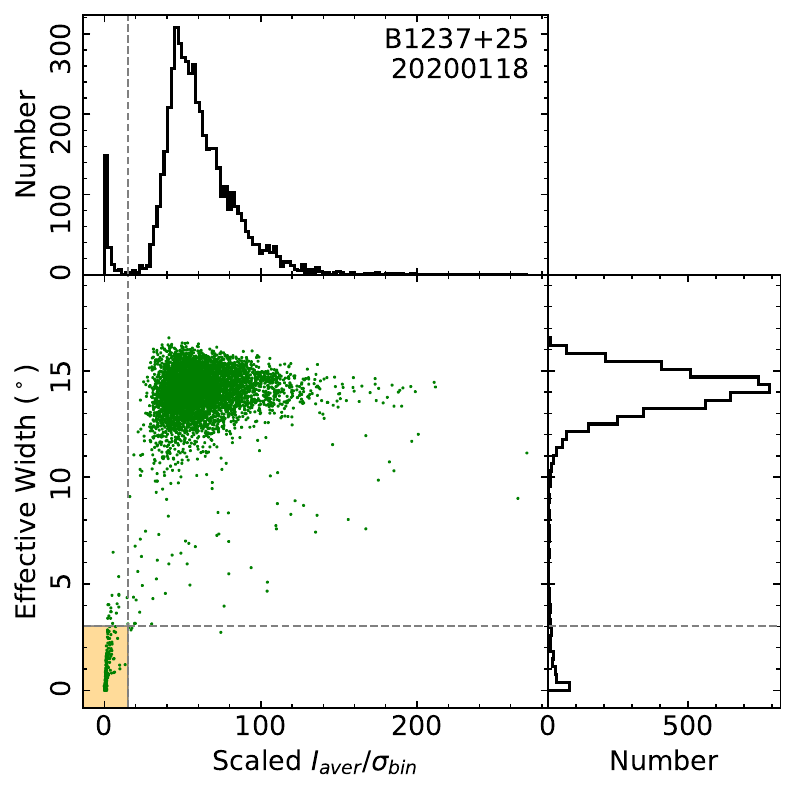}
\end{tabular}
\caption{The same as Figure~\ref{FigIW_J0528_20220304} but for the effective pulse width and pulse intensity for single pulses of PSR B1237+25 observed by FAST on 2020 January 18.}
\label{FigIW_J1239}
\end{figure}

\begin{figure}
\centering
\setlength\tabcolsep{0pt}
\begin{tabular}{ccc}
\includegraphics[width=0.42\textwidth]{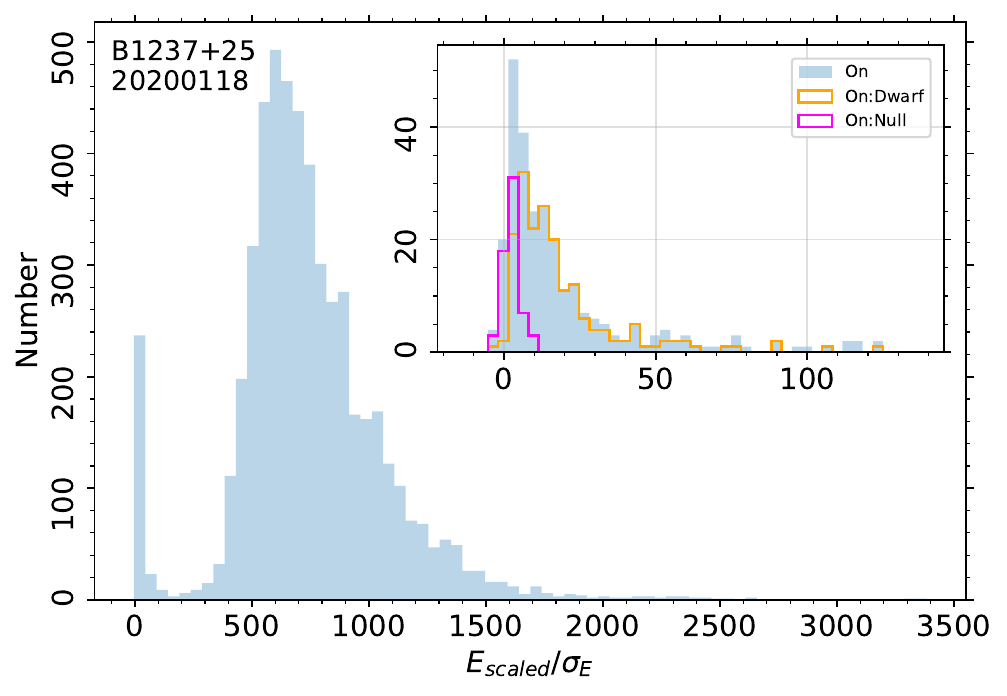}
\end{tabular}
\caption{The same as Figure~\ref{FigIhist_J0528_20220304} but for pulse energy of PSR B1237+25 observed by FAST on 2020 January 18.}
\label{FigIhist_J1239}
\end{figure}

\begin{figure}
\centering
\includegraphics[width=0.4\textwidth]{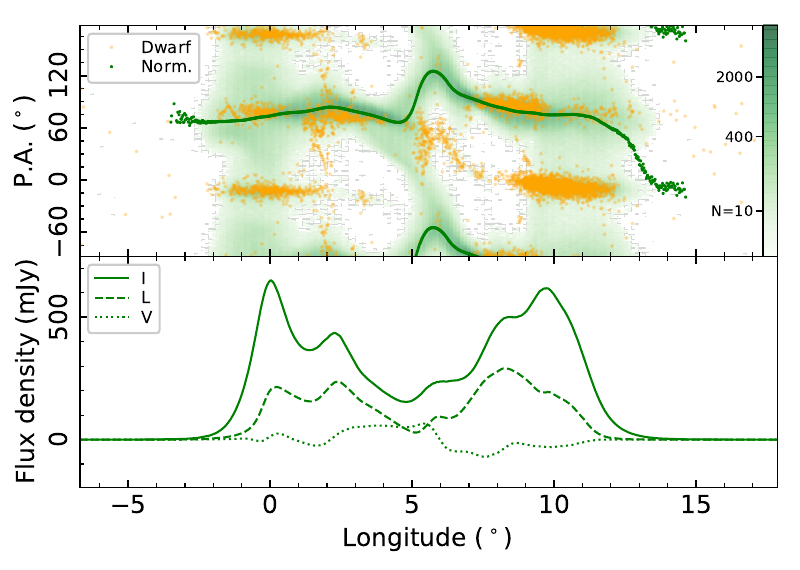}
\caption{The same as Figure~\ref{FigPolCon_J0528_20220304} but for  dwarf pulses and normal pulses of PSR B1237+25 observed by FAST on 2020 January 18. }
\label{FigPolCon_J1239}
\end{figure}

%


\subsection{PSR B1237+25}

PSR B1237+25 is a bright pulsar discovered by \citet{Lang1969}. 
The pulsar exhibits four states, nulling state \citep{Backer1970_Nulling}, ``abnormal" mode, ``quiet-normal" mode with 2.8-period modulation, and ``flare-normal" mode with a modulation of about four periods \citep{Backer1970_Moding,Backer1970_modulation,Srostlik2005}. This pulsar has five main components, with the central one from the core emission and other two pairs from the inner cone and outer cone \citep[e.g.][]{,Timothy2019}. 
Very weak emission is previously detected by \citet{Srostlik2005} during nulls.

We process FAST achieve data for PSR B1237+25 observed for 2 hr on 2020 January 18, and  detect 191 dwarf pulses (see Figure~\ref{FigSinPulStack_J1239} and Table~\ref{table:obs}). To diminish the modulation from scintillation, the profiles are scaled by the averaged intensity of nearby pulses. The distributions of the effective pulse width and averaged intensity are used to distinguish dwarf pulses and normal pulses, as shown in Figure~\ref{FigIW_J1239}. The effective emission width is counted by bins with an intensity exceeding 5$\sigma_{\rm bin}$.

Similar to that for PSR B2111+46 in \citet{Chen2023}, the dwarf pulses are obviously a very distinct population in the left-bottom corner of Figure~\ref{FigIW_J1239}, with a narrower effective emission width of less than 3$^\circ$ and a lower averaged intensity of less than 15$\sigma_{\rm bin}$. The energy histogram of individual pulses of PSR B1237+25 is clearly bimodal (see Figure~\ref{FigIhist_J1239}). The peak around zero intensity is contributed by dwarf pulses and ``nulls". Therefore, ``nulls" and dwarf pulses come from the same population according to the distribution. Notably, the energy distribution of ``nulls" is asymmetric of zero energy, indicating that the null state may be the very weak single-pulse emission even not well detected by the FAST.
%
%
Details of dwarf pulses of PSR B1237+25 are also intriguing, as depicted by one example in Figure~\ref{FigSinPulStack_J1239}. Some dwarf pulses are isolated and very narrow, spanning only one or two bins, such as the subpulse for the individual pulse No. 4138.

As shown in Figure~\ref{FigPolCon_J1239}, the dwarf pulses of PSR B1237+25 can appear in any emission longitudes within the emission window defined by the mean profile of both the core and two conal emission components \citep{Srostlik2005, Young2012, Timothy2019}. 
Polarization properties of dwarf pulses are somewhat different from normal pulses in some aspects. For those from the two inner cone components, the PA distribution of dwarf pulses are consistent with that of normal pulses, but those from the core and outer cone components are mostly in the orthogonal polarization mode. 

For PSR B1237+25, we do detect narrow bright pulses in the core region of mean profile, 
and these pulses are not so ``dwarf'', and therefore are not included in the analyses for the dwarf pulses.

\begin{figure*}
\centering
\renewcommand\arraystretch{0.01}
\begin{minipage}{0.28\linewidth}
    \centering
    \includegraphics[width=0.99\linewidth]{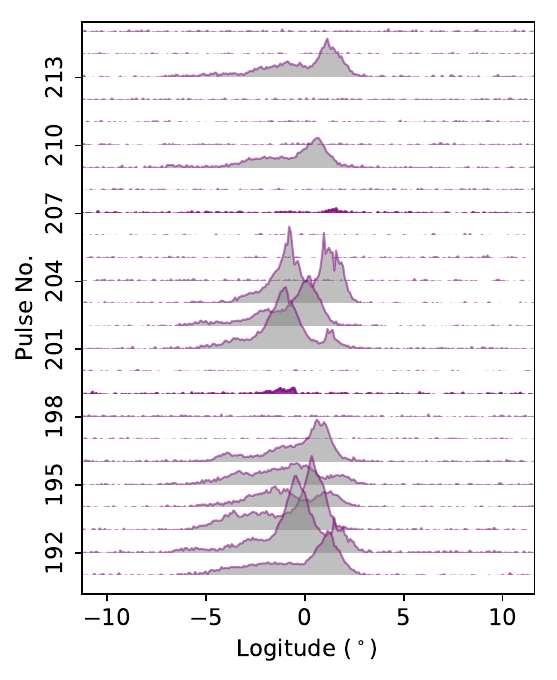}
\end{minipage}
\begin{minipage}{0.64\linewidth}
    \centering
    \includegraphics[width=0.99\linewidth]{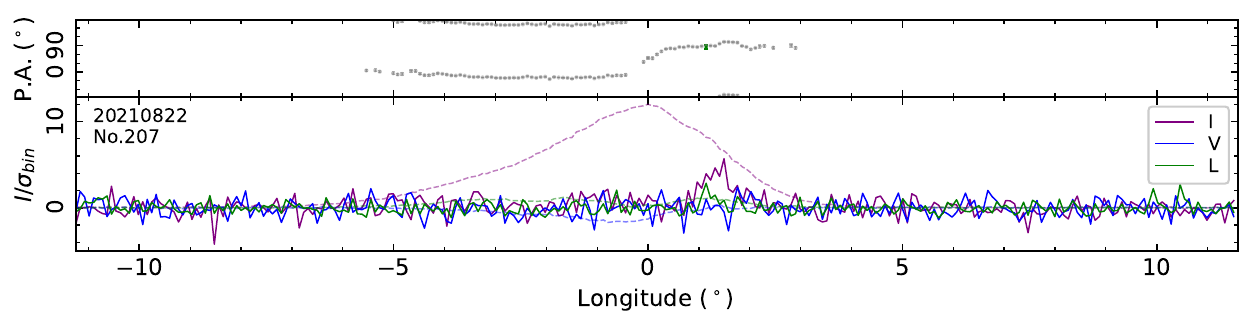}
    \includegraphics[width=0.99\linewidth]{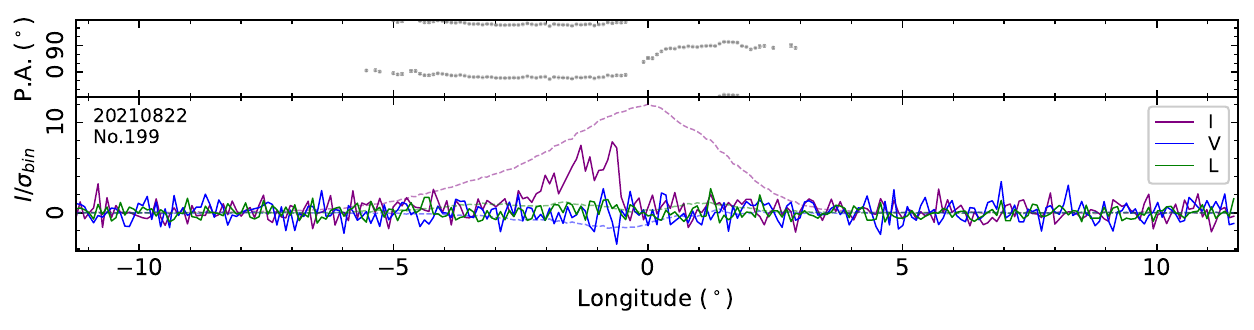}
\end{minipage}
\caption{The same as Figure~\ref{FigSinPulStack_J0528_20220304}, but for two dwarf pulses of PSR J1851$-$0053 observed on 2021 August 22. Polarization profiles have a time resolution of 0.34 ms, i.e. 4096 bins per period.}
\label{FigSinPulStack_J1851_20210822}
\end{figure*}

\begin{figure}
\centering
\includegraphics[width=0.35\textwidth]{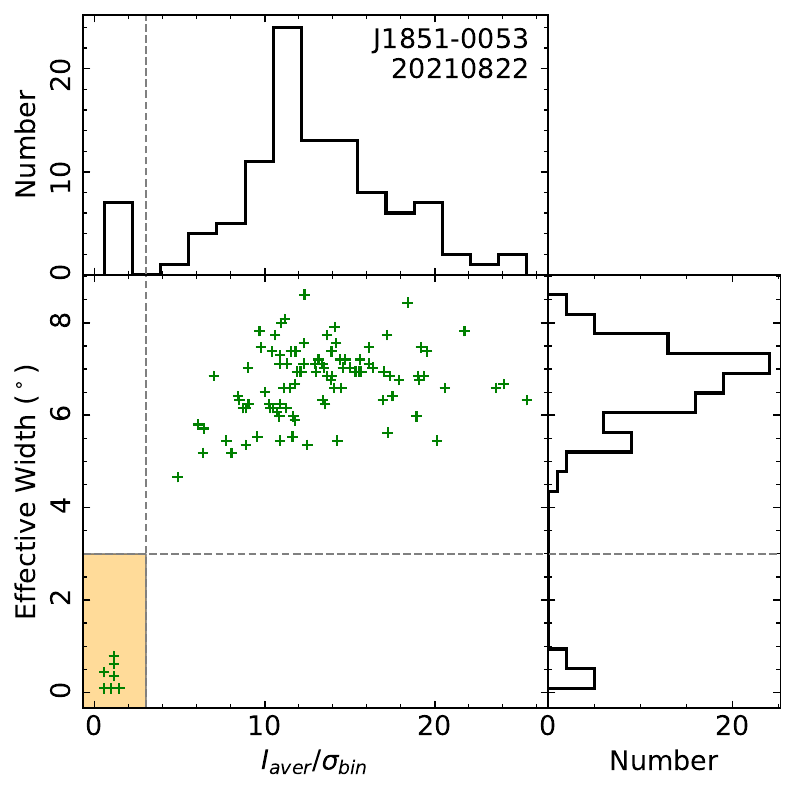}
\caption{The same as Figure~\ref{FigIW_J0528_20220304} but for the effective pulse width and the averaged intensity for single pulses of PSR J1851-0053 observed on 2021 August 22.}
\label{FigIW_J1851_20210822}
\end{figure}

\begin{figure*}
\centering
\begin{minipage}{0.275\linewidth}
    \centering
    \includegraphics[width=0.99\linewidth]{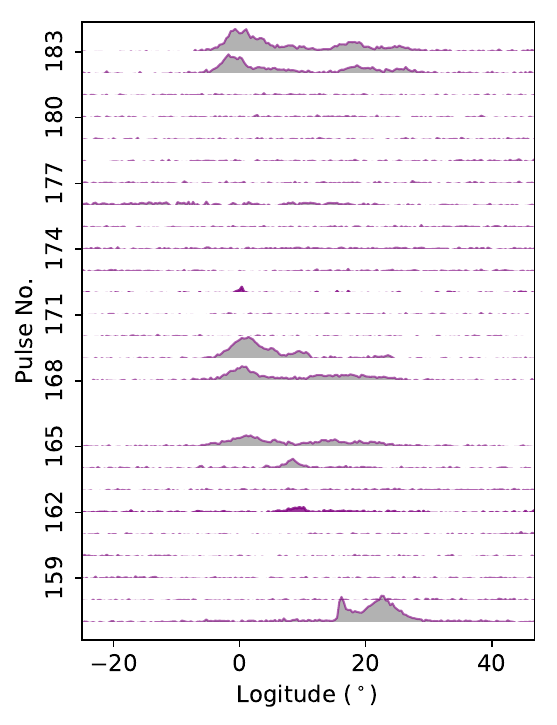}
\end{minipage}
\begin{minipage}{0.68\linewidth}
    \centering
    \includegraphics[width=0.99\linewidth]{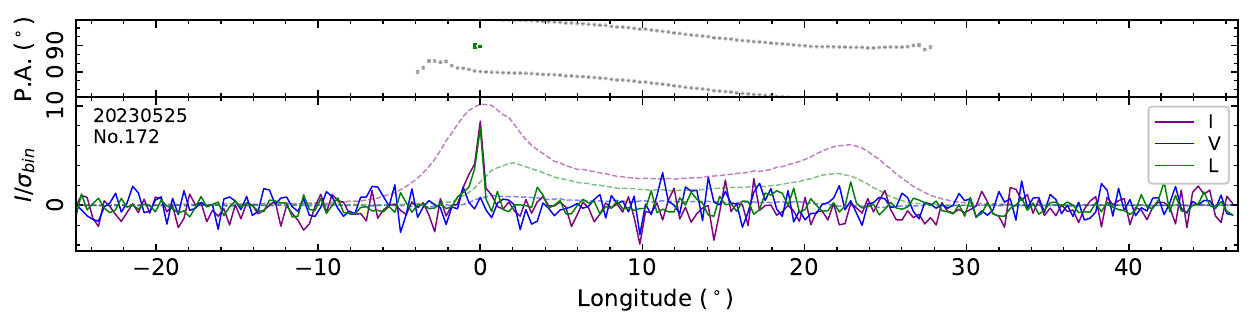}
    \includegraphics[width=0.99\linewidth]{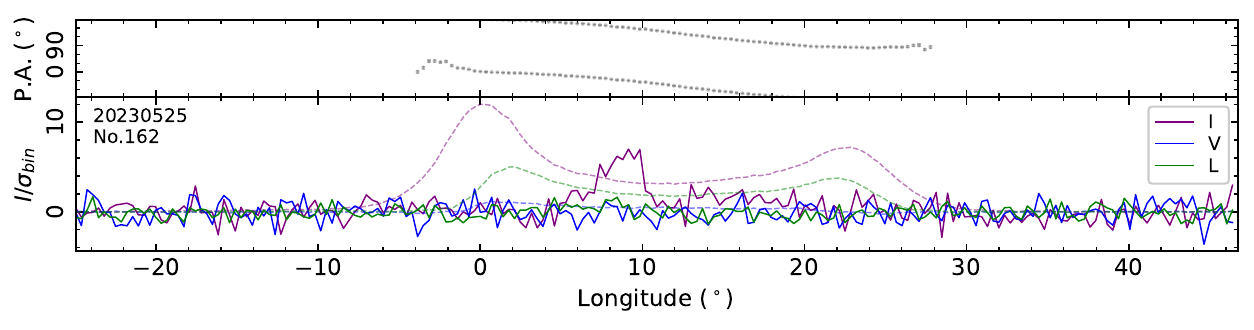}
\end{minipage}
\caption{The same as Figure~\ref{FigSinPulStack_J0528_20220304}, but for two dwarf pulses of PSR B1901+10 observed on 2023 May 25. Polarization profiles have a time resolution of 1.81 ms, i.e. 1024 bins per period.
}
\label{FigSinPulStack_J1904_20230525_0}
\end{figure*}

\begin{figure}  \centering
\includegraphics[width=0.35\textwidth]{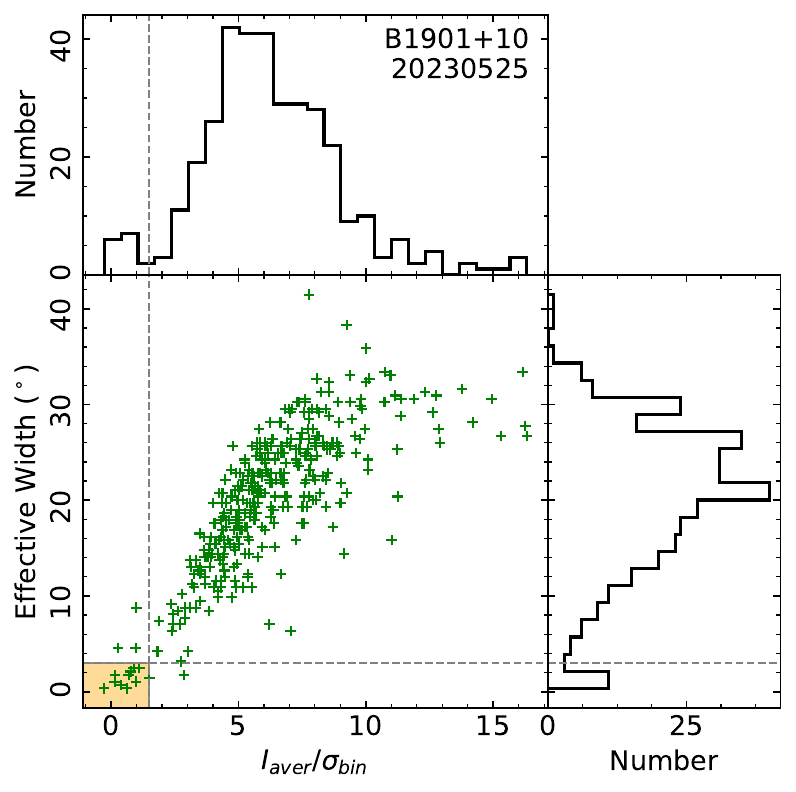}
\caption{
The same as Figure~\ref{FigIW_J0528_20220304} but for the effective pulse width and the averaged intensity for single pulses of PSR B1901+10 observed by FAST on 2023 May 25. }
\label{FigIW_J1904_20230525}
\end{figure}

\begin{figure}
\centering
\setlength\tabcolsep{0pt}
\begin{tabular}{ccc}
\includegraphics[width=0.4\textwidth]{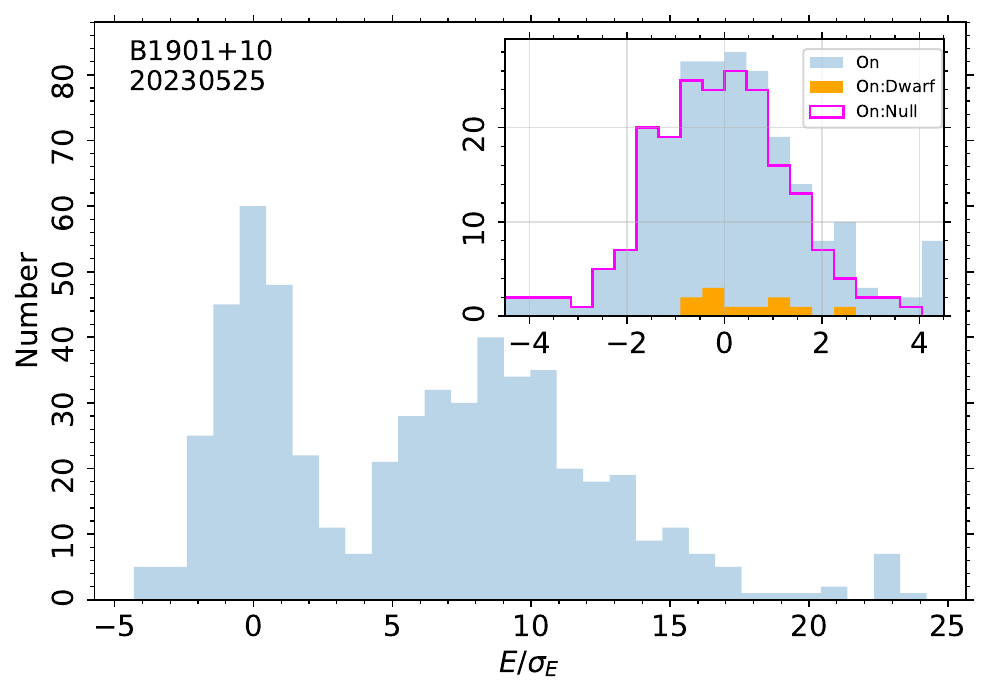}
\end{tabular}
\caption{The same as Figure~\ref{FigIhist_J0528_20220304} but for pulse energy of PSR B1901+10 observed by FAST on 2023 May 25.}
\label{FigIhist_J1904_20230525}
\end{figure}



\begin{figure*} 
\centering
\begin{minipage}{0.29\linewidth}
    \centering
    \includegraphics[width=0.99\linewidth]{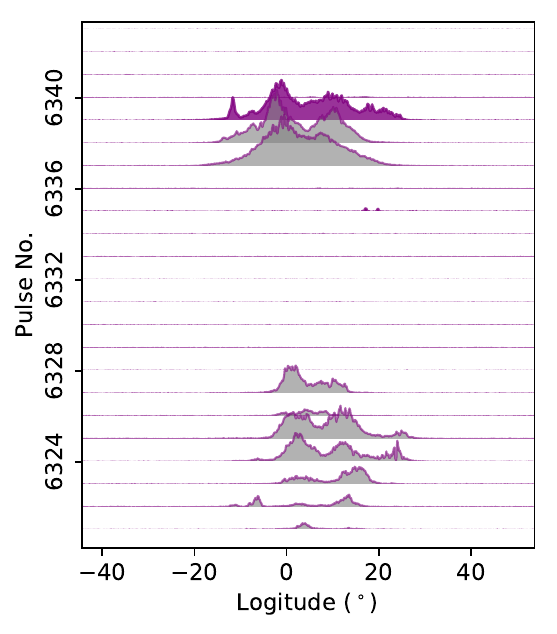}
\end{minipage}
\begin{minipage}{0.64\linewidth}
    \centering
    \includegraphics[width=0.99\linewidth]{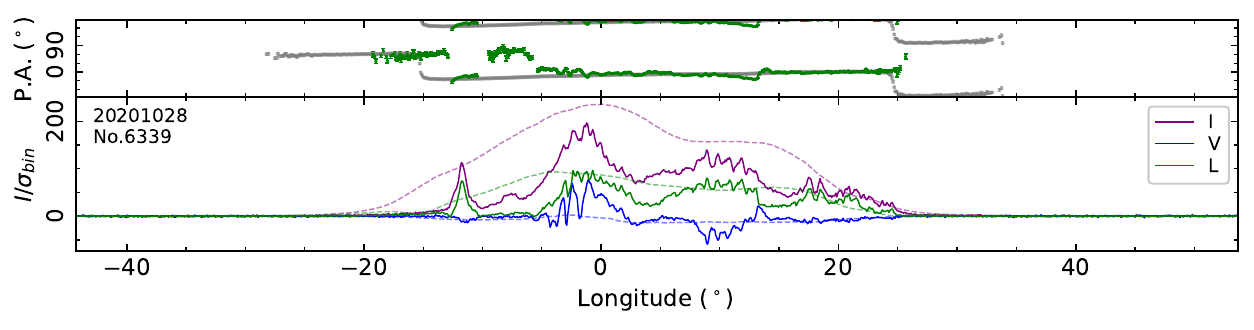}
    \includegraphics[width=0.99\linewidth]{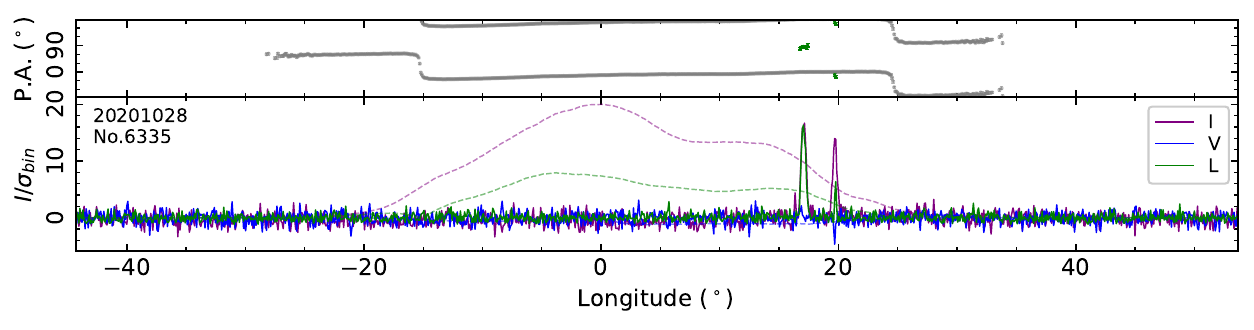}
\end{minipage}
\caption{
The same as Figure~\ref{FigSinPulStack_J0528_20220304}, but for a normal pulse and a dwarf pulse of PSR B1944+17 observed on 2020 October 28. Polarization profiles have a time resolution of 0.11 ms, i.e. 4096 bins per period.
}
\label{FigSinPulStack_J1946_20201028}
\end{figure*}

\begin{figure}
\centering
\includegraphics[width=0.38\textwidth]{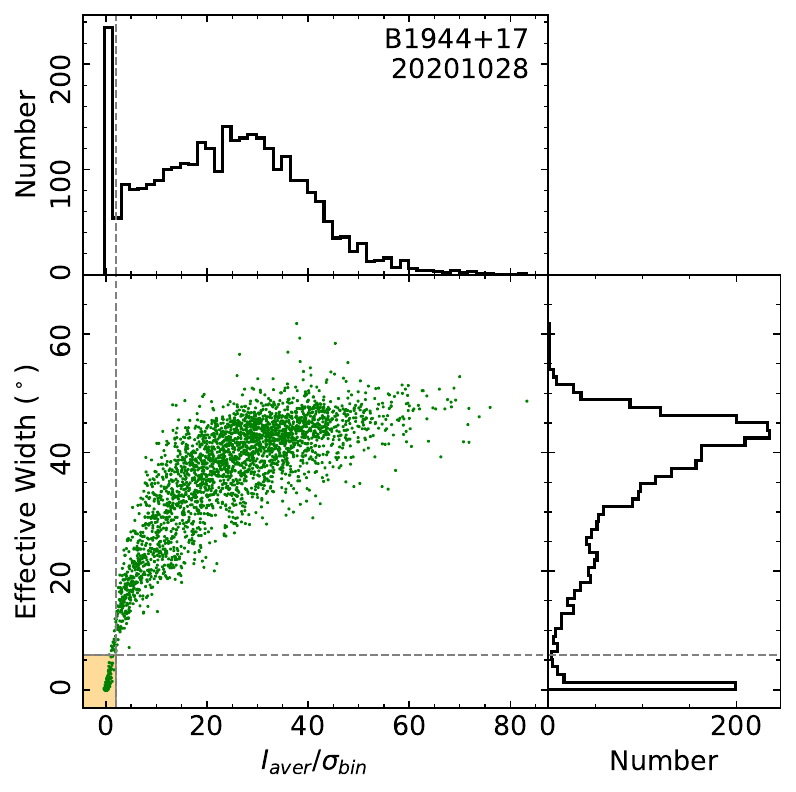}
\includegraphics[width=0.38\textwidth]{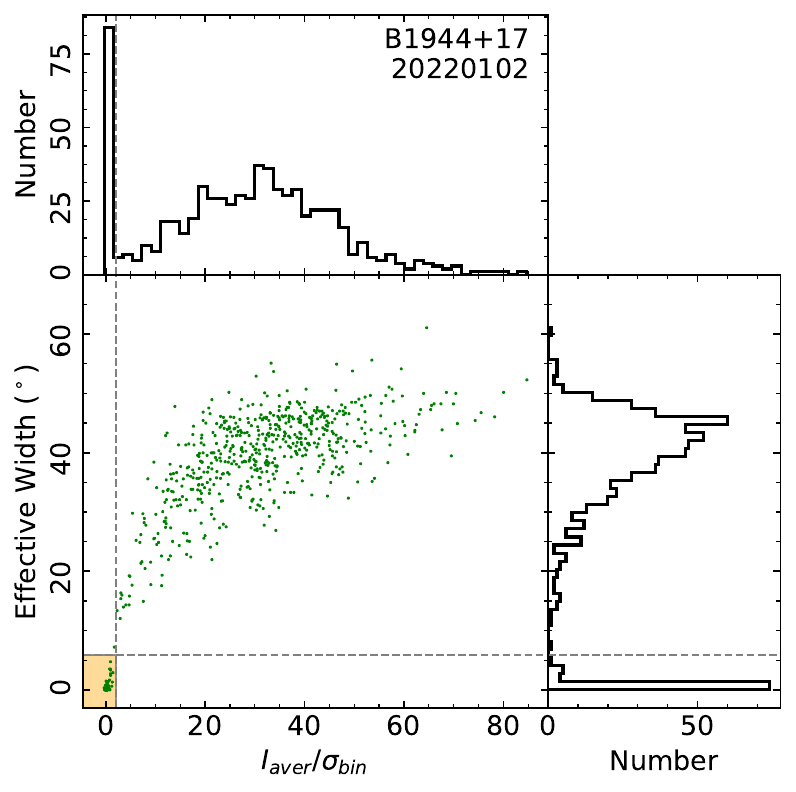}
\caption{
The same as Figure~\ref{FigIW_J0528_20220304} but for the effective pulse width and the averaged intensity for single pulses of PSR B1944+17 observed by FAST on 2020 October 28 and 2022 January 2. Dwarf pulses are a distinct group from normal pulses.}
\label{FigIW_J1946_2dates}
\end{figure}

\begin{figure}
\centering
\includegraphics[width=0.4\textwidth]{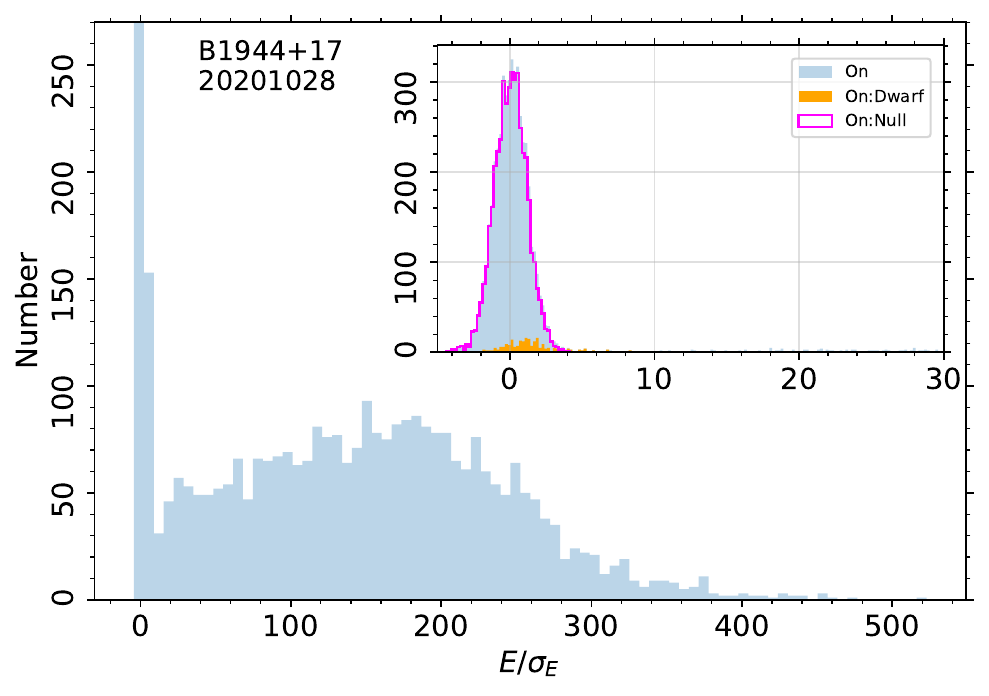}
\caption{The same as Figure~\ref{FigIhist_J0528_20220304} but for pulse energy of PSR B1944+17 observed by FAST on 2020 October 28. }
\label{FigIhist_J1946_20201028}
\end{figure}

\begin{figure}
\centering
\includegraphics[width=0.4\textwidth]{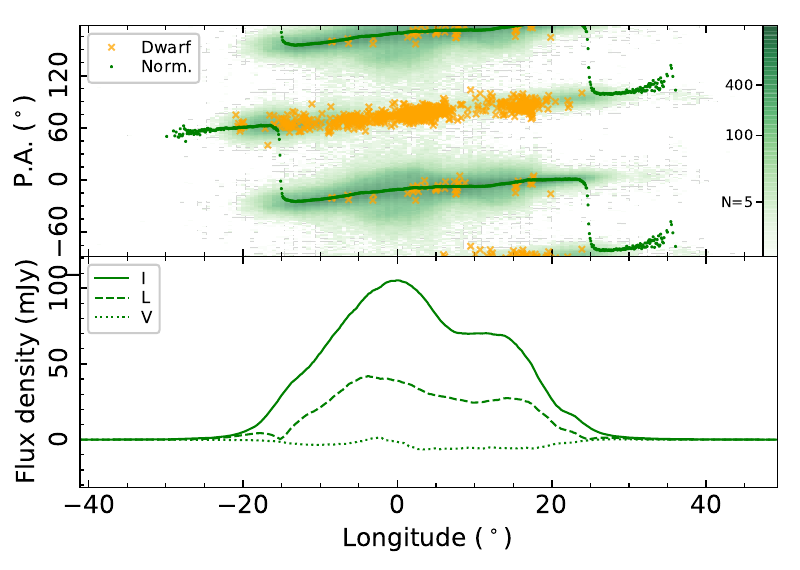}
\caption{ The same as Figure~\ref{FigPolCon_J0528_20220304} but for dwarf pulses and normal pulses of PSR B1944+17 observed on date 2020 October 28.}
\label{FigPolCon_J1946_20201028}
\end{figure}

\begin{figure}
\centering
\includegraphics[width=0.46\textwidth]{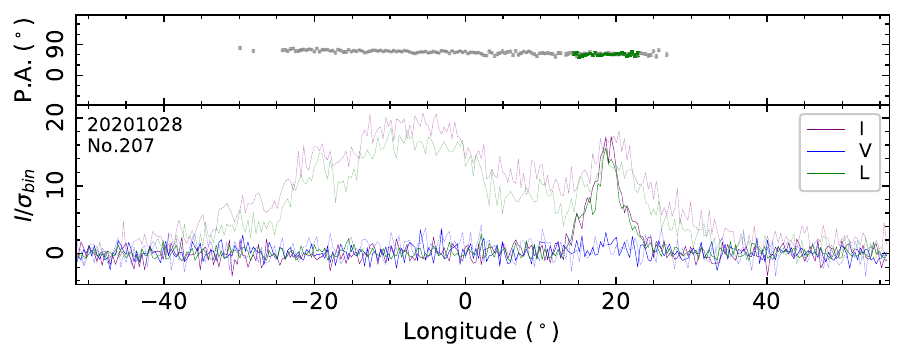}
\includegraphics[width=0.46\textwidth]{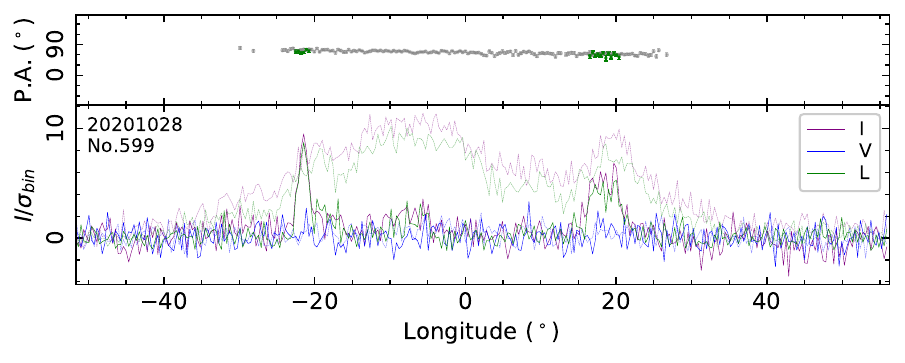}
\caption{Two examples for weak single pulses in the longitude range of the interpulse for PSR B1944+17 observed on 2020 October 28. The mean polarization profiles of the interpulse are plotted in the background, and all data have a time resolution of 0.43 ms, i.e. 1024 bins per period.}
\label{FigIP_J1946_20201028}
\end{figure}

\subsection{PSR J1851$-$0053}

PSR J1851$-$0053 is a southern pulsar discovered  by the Parkes telescope during the multibeam pulsar survey \citep{Hobbs2004}.
It was observed by chance during a FAST GPPS survey observation on 2021 August 22 for 5 minutes. We detect seven dwarf pulses from some nulling periods.  See Figure~\ref{FigSinPulStack_J1851_20210822} for two examples. The distribution of the effective width and the averaged intensity is shown in Figure~\ref{FigIW_J1851_20210822}, and this distinct group of dwarf pulses has an effective width of less than 3$^\circ$ and an averaged intensity of less than 3$\sigma_{\rm bin}$.


\subsection{PSR B1901+10}

PSR B1901+10 is a bright pulsar discovered by Arecibo \citep{Hulse1975} and is known for nulling \citep{Lorimer2002}. This pulsar was observed for 18 minutes by FAST on 2023 May 25 in an applied project by the first author to study narrow pulses of nulling pulsars. As depicted in Figure~\ref{FigSinPulStack_J1904_20230525_0}, individual pulses often exhibit nulling behavior. 

The sensitive FAST observation detects individual pulses with varying pulse width and intensity, and gives the distribution of pulse intensity and the effective width of single pulses in Figure~\ref{FigIW_J1904_20230525}. The 11 dwarf pulses belong to a distinct group with an intensity of less than 1.5$\sigma_{\rm bin}$ and a width of less than 3$^\circ$. Because of the low signal-to-noise ratio, these dwarf pulses are hidden in the nulling phase (see Figure~\ref{FigIhist_J1904_20230525}).

\subsection{PSR B1944+17}

PSR B1944+17 is a bright pulsar \citep{Vaughan1970} which also has nulling, drifting and mode-changing phenomena \citep{Deich1986} and has four emission modes \citep{Kloumann2010}. About two-thirds of periods are nulling, some nulls are short, and some nulls last for many periods. \citet{Kloumann2010} found that the integration of short nulls of less than eight periods gives a weak emission, but that for longer nulls there is no measurable emission. They suggest that short nulls are produced by the sight line passing through the empty area of rotating ``carousels" of subbeams, and long nulls are caused by emission cessation.

We got PSR B1944+17 observed for 15 minutes during a verification observation on 2022 January 2 for a pulsar candidate in the FAST GPPS survey, and detect 84 dwarf pulses. In order to study the nature of dwarf pulses, we process archived data from a FAST observation conducted on 2020 October 28 for 55 minutes, and identify 234 dwarf pulses (see Figures~\ref{FigSinPulStack_J1946_20201028} and~\ref{FigIW_J1946_2dates}). The statistic results on the dwarf pulses from these two observations are consistent, and longer observation provides a better concentration of data distributions. The distributions of effective width and averaged intensity for individual pulses from the two observations are shown in Figure~\ref{FigIW_J1946_2dates}. The dwarf pulses are obviously in a distinct group of a low average intensity of less than 2$\sigma_{\rm bin}$  and a narrow emission width of less than 5.$^\circ$9. Again, as seen in Figure~\ref{FigIhist_J1946_20201028}, these dwarf pulses have a small energy hidden in the histogram peak for nulling periods. 

Some dwarf pulses have only one narrow weak subpulse, but some are composed of two or more discrete subpulses. They can appear in any period in either short- or long-duration nullings, suggesting that the nulls with a long duration are also not the case of real emission cessation,  in contrast to \citet{Kloumann2010}.

Figure~\ref{FigPolCon_J1946_20201028} shows the averaged polarization profile of 2828 normal pulses, as well as the PA distribution of normal pulses and the dwarf pulses. The  dwarf pulses are more likely to appear in the central part of profile components, corresponding to the strongest component in the averaged profile.
Polarization properties of normal pulses and dwarf pulses are very distinct. Most of normal pulses in the longitude range from -15$^\circ$ to 25$^\circ$ have one polarization mode, while most dwarf pulses in this longitude range are on the orthogonal polarization mode, the same mode for the normal pulses in the two edges beyond the phase range. 

This pulsar has a very weak interpulse in the mean profile, as identified by \citet{Kloumann2010} which is suggested to be core emission. Based on the highly sensitive observations by FAST, we have detected the interpulse which has two or three profile components (see Figure~\ref{FigIP_J1946_20201028}), all highly polarized. We also detect some individual subpulses in the longitude range of the interpulse with a good signal-to-noise ratio, and they have a pulse width and intensity similar to dwarf pulses in the main pulses. They are highly polarized, as seen for examples in Figure~\ref{FigIP_J1946_20201028}. The high linear polarization and not significant circular polarization are indications of the conal emission in nature, rather than the core emission.



\subsection{Five pulsars with few dwarf pulses detected}

\begin{figure*}
\centering
\renewcommand\arraystretch{0.01}
\begin{minipage}{0.29\linewidth}
    \centering
    \includegraphics[width=0.99\linewidth]{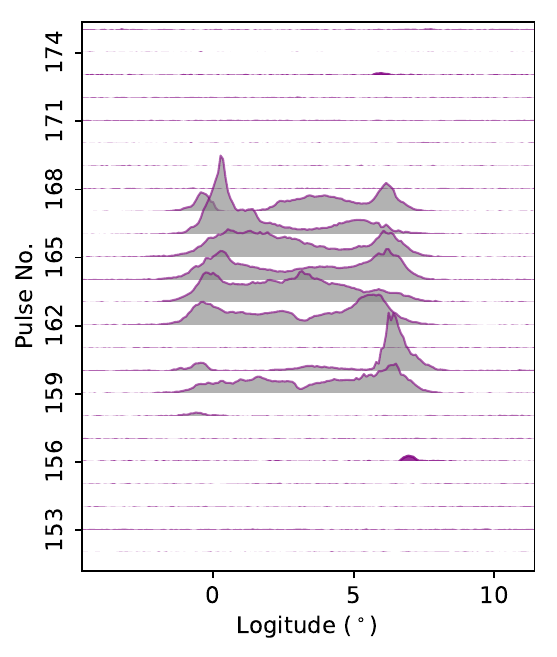}
\end{minipage}
\begin{minipage}{0.64\linewidth}
    \centering
    \includegraphics[width=0.99\linewidth]{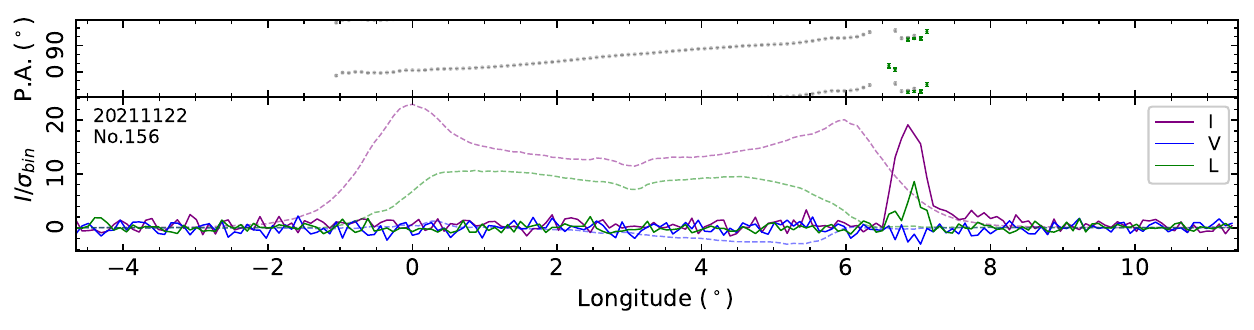}
    \includegraphics[width=0.99\linewidth]{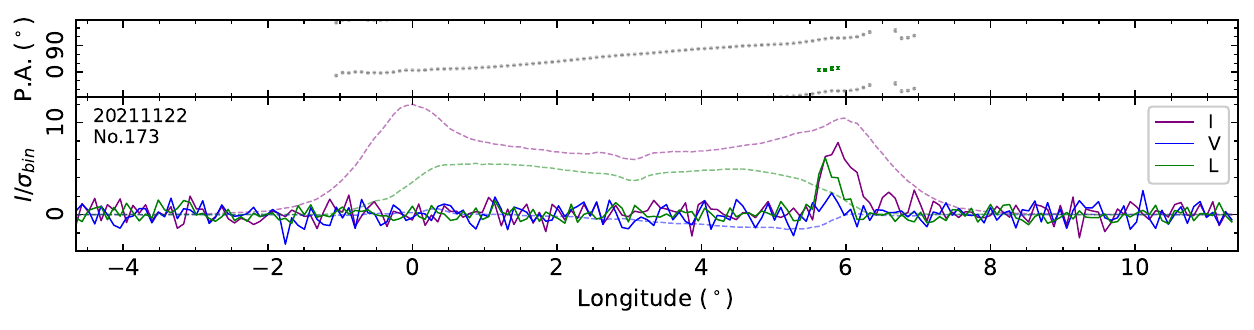}
\end{minipage}
\caption{The same as Figure~\ref{FigSinPulStack_J0528_20220304}, but for two dwarf pulses of PSR J1538+2345 observed on 2021 November 22. Polarization profiles have a time resolution of 0.84 ms, i.e. 4096 bins per period.}
\label{FigSinPulStack_J1538_20211122}
\end{figure*}

\begin{figure*}
\centering
\renewcommand\arraystretch{0.01}
\begin{minipage}{0.3\linewidth}
    \centering
    \includegraphics[width=0.99\linewidth]{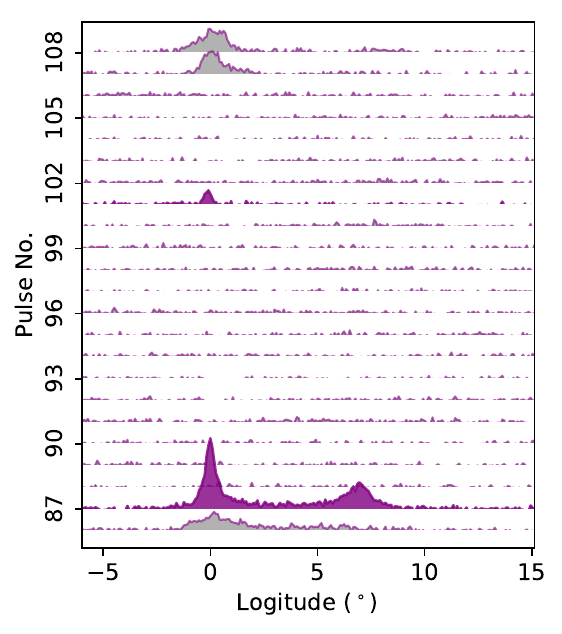}
\end{minipage}
\begin{minipage}{0.64\linewidth}
    \centering
    \includegraphics[width=0.99\linewidth]{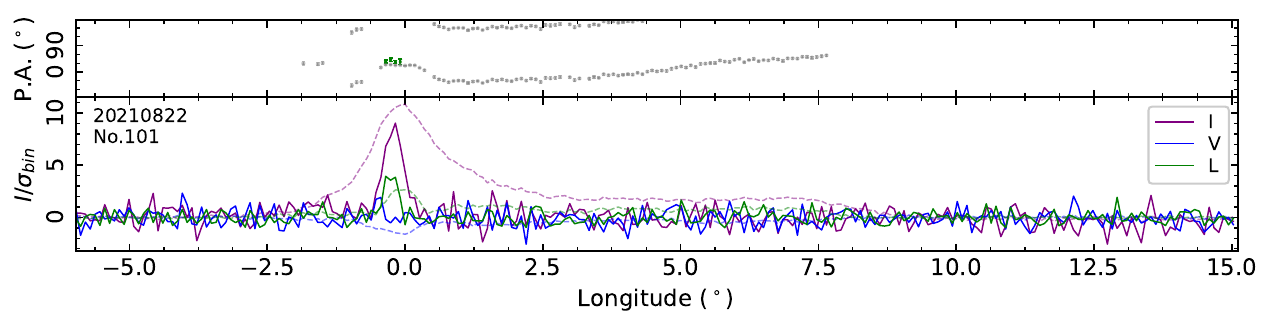}
    \includegraphics[width=0.99\linewidth]{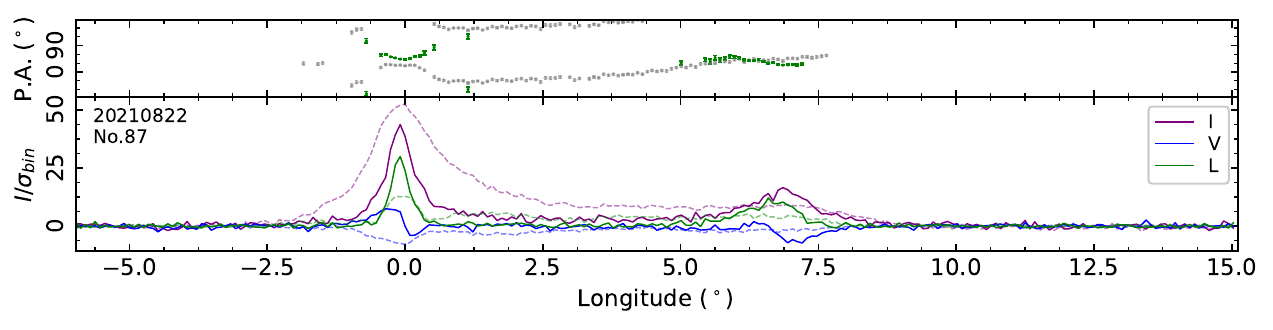}
\end{minipage}
\caption{The same as Figure~\ref{FigSinPulStack_J0528_20220304}, but for a single dwarf pulse and a normal pulse of PSR J1824-0127 observed on 2021 August 22. Polarization profiles have a time resolution of 0.61 ms, i.e. 4096 bins per period. }
\label{FigSinPulStack_J1824}
\end{figure*}

\begin{figure*}
\centering
\begin{minipage}{0.29\linewidth}
    \centering
    \includegraphics[width=0.99\linewidth]{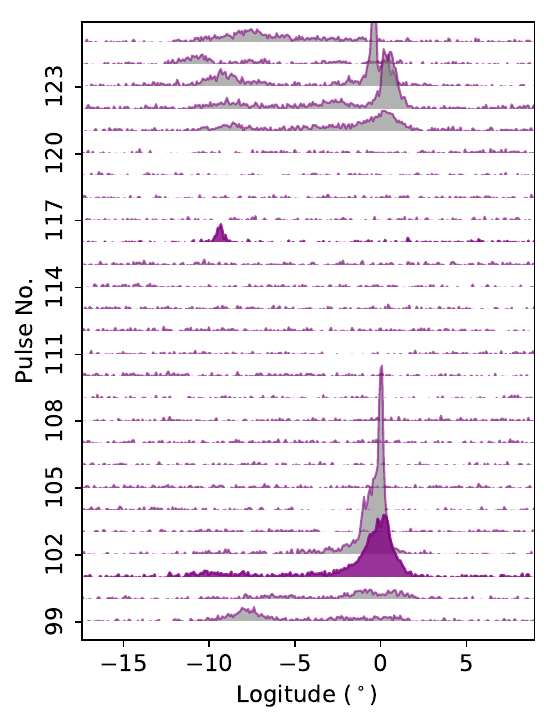}
\end{minipage}
\begin{minipage}{0.64\linewidth}
    \centering
    \includegraphics[width=0.99\linewidth]{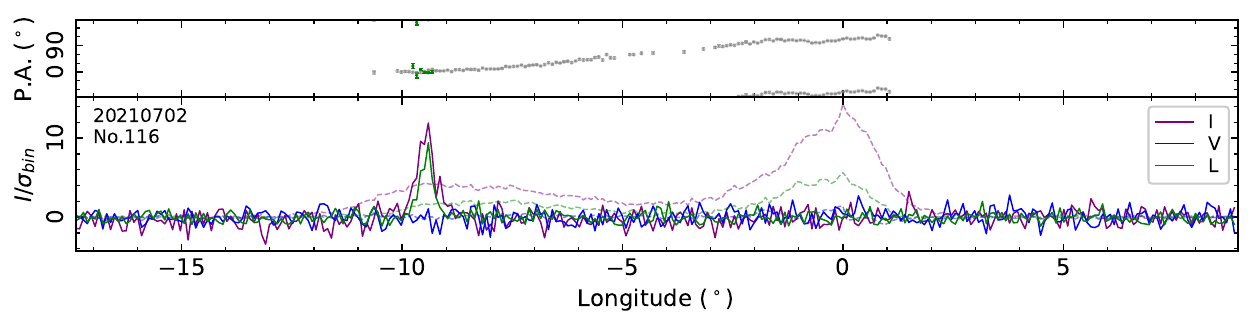}
    \includegraphics[width=0.99\linewidth]{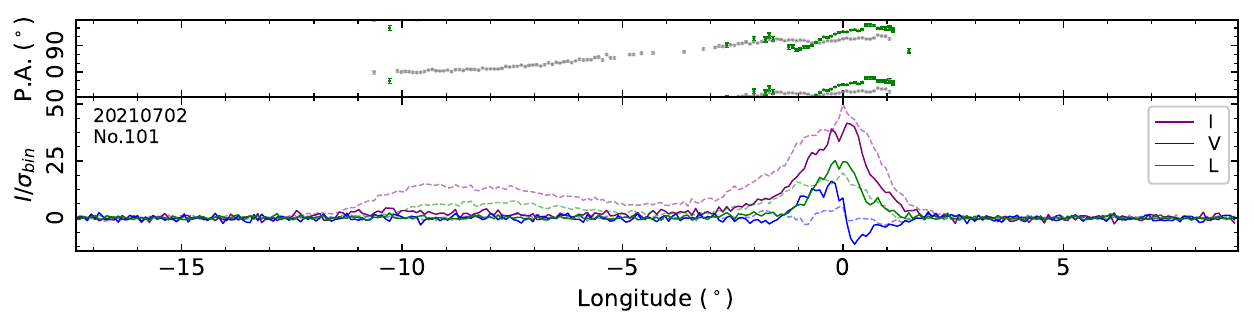}
\end{minipage}
\caption{The same as Figure~\ref{FigSinPulStack_J0528_20220304}, but for a dwarf pulse and a normal pulse of PSR J1939+10  observed on 2021 July 2. Polarization profiles have a time resolution of 0.56 ms, i.e. 4096 bins per period.
}
\label{FigSinPulStack_J1939_20210702}
\end{figure*}

\begin{figure*}
\centering
\setlength\tabcolsep{0pt}
\renewcommand\arraystretch{0.01}
\begin{minipage}{0.29\linewidth}
    \centering
    \includegraphics[width=0.99\linewidth]{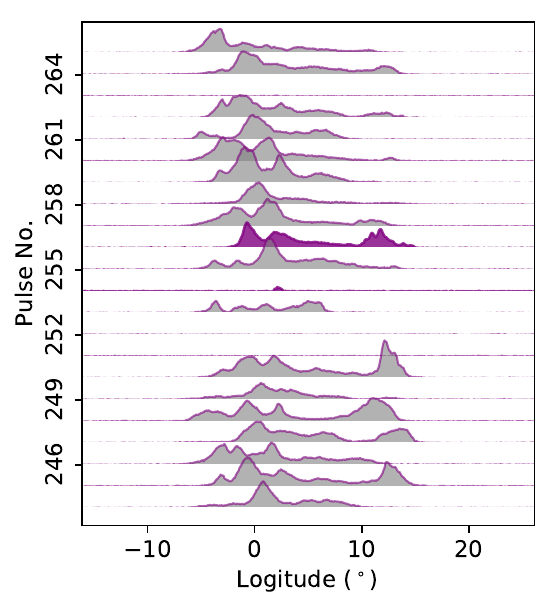}
\end{minipage}
\begin{minipage}{0.64\linewidth}
    \centering
    \includegraphics[width=0.99\linewidth]{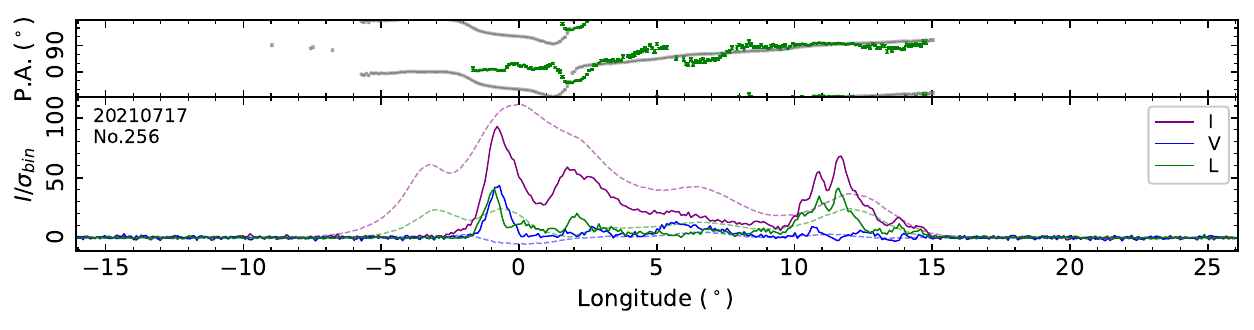}
    \includegraphics[width=0.99\linewidth]{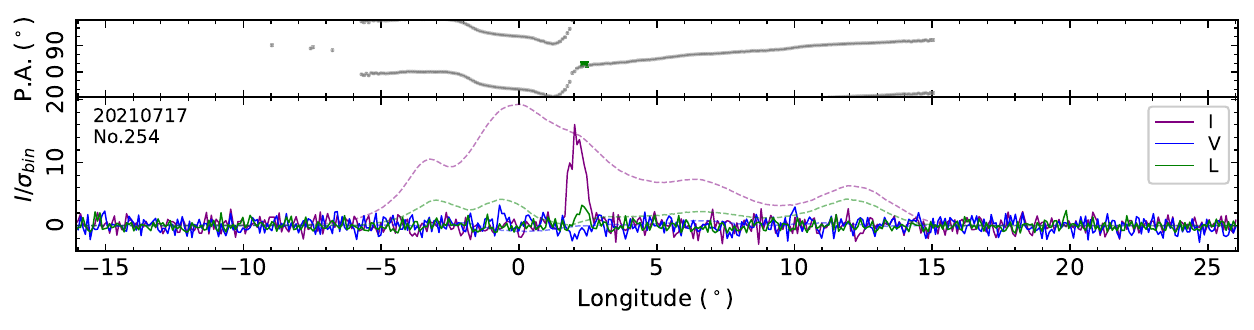}
\end{minipage}
\caption{The same as Figure~\ref{FigSinPulStack_J0528_20220304}, but for a dwarf pulse and a normal pulse of PSR B2000+40 observed on 2021 July 17. Polarization profiles have a time resolution of 0.22 ms, i.e. 4096 bins per period.}
\label{FigSinPulStack_J2002_20210717}
\end{figure*}

\begin{figure*}
\centering
\setlength\tabcolsep{0pt}
\renewcommand\arraystretch{0.01}
\begin{minipage}{0.29\linewidth}
    \centering
    \includegraphics[width=0.99\linewidth]{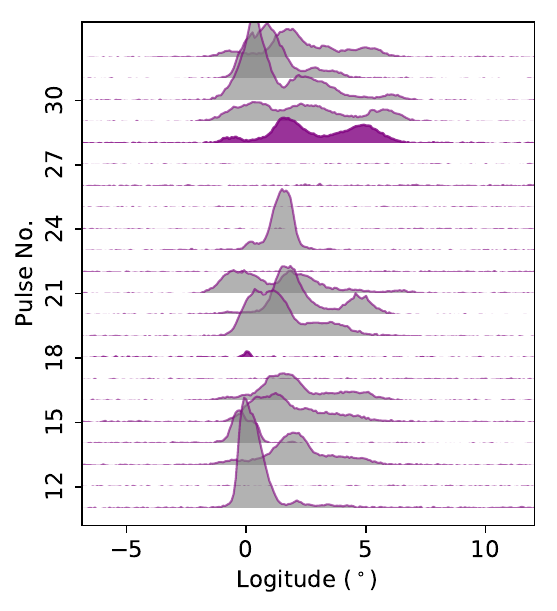}
\end{minipage}
\begin{minipage}{0.64\linewidth}
    \centering
    \includegraphics[width=0.99\linewidth]{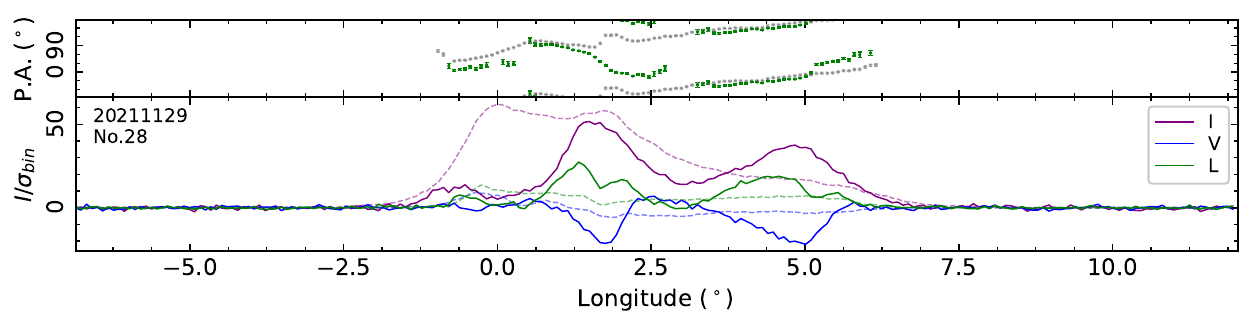}
    \includegraphics[width=0.99\linewidth]{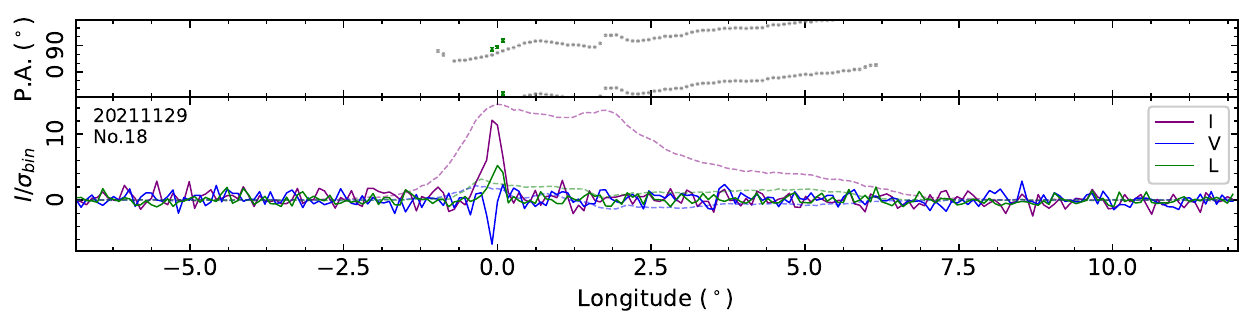}
\end{minipage}
\caption{The same as Figure~\ref{FigSinPulStack_J0528_20220304}, but for a dwarf pulse and a normal pulse of PSR J2112+4058 observed on 2021 November 29. Polarization profiles have a time resolution of 0.99 ms, i.e. 4096 bins per period.}
\label{FigSinPulStack_J2112_20211129}
\end{figure*}

\begin{figure}
\centering
\includegraphics[width=0.43\textwidth]{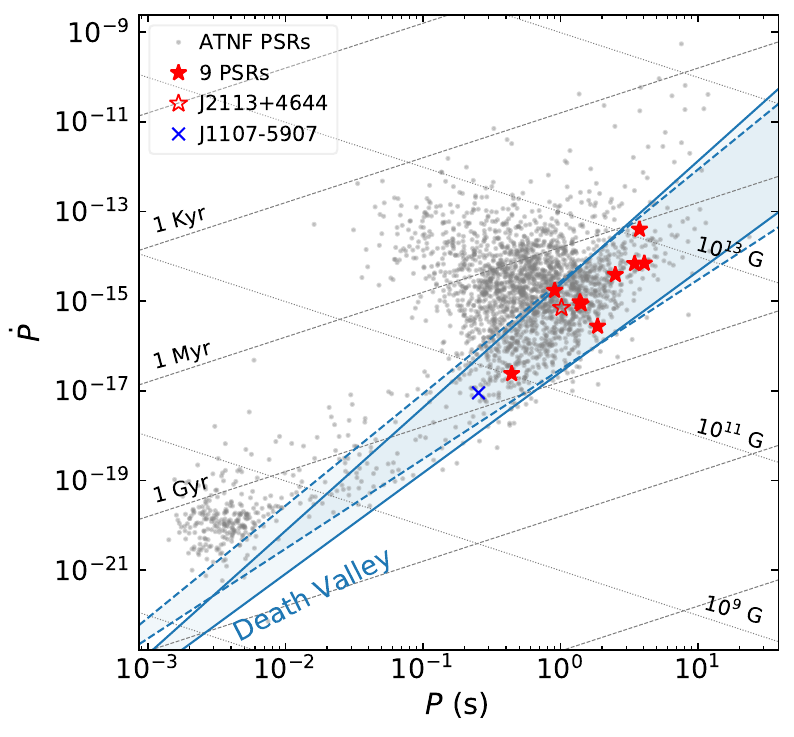}
\caption{Nine pulsars with dwarf pulses detected in this paper (note that PSR J1939+10 is not included for no period derivative available; two pulsars are almost overlapped) are located in the the death valley  \citep{Zhang2000} in the $P-\dot{P}$ diagram. Pulsar data are taken from the pulsar catalog \citep[gray;][]{Manchester2005}. PSRs J2113+4644 \citep{Chen2023} and J1107-5907 \citep{Young2014} are also marked. }
\label{FigPPdot}
\end{figure}

\subsubsection{PSR J1538+2345}

PSR J1538+2345 was discovered as a RRAT at 350 MHz in the drift-scan survey of the Green Bank Telescope \citep{Karako2015}. \citet{Zhou2023} reported the source to be just a pulsar with nulling features in the FAST observation. Using this 15 minute data on 2021 November 22, we detected four dwarf pulses. A dwarf pulse is displayed in  Figure~\ref{FigSinPulStack_J1538_20211122}. We also processed the released archive data observed by FAST for 10 minutes on 2020 December 16, and another two dwarf pulses are detected.

\subsubsection{PSR J1824$-$0127}

PSR J1824$-$0127 is a southern pulsar discovered by the Parkes telescope during the multibeam pulsar survey \citep{Lorimer2006}. 
We get two FAST observations for this pulsar. The 5 minute FAST GPPS survey observation was made on 2021 August 22, and the 15 minute tracking observation was made on 2022 September 2 during a verification of a pulsar candidate. We detect the nulling phenomenon of this pulsar for the first time, and find three dwarf pulses during the nulling phases. One example is shown in Figure~\ref{FigSinPulStack_J1824}.

\subsubsection{PSR J1939+10}

PSR J1939+10 was discovered by the Arecibo telescope \citep{Camilo1996}. The exact position is not known, probably because of lack of timing observations. We detect this pulsar by chance during a 5 minute session of the FAST GPPS survey on 2021 July 2, and we get a coarse position R.A.=19:39:22.6 and decl.=+10:49:54, with a position uncertainty of 1.5 arcminutes \citep{Han2021}. 
We detect only one dwarf pulse (see Figure~\ref{FigSinPulStack_J1939_20210702}) during a nulling period, which in fact has a very narrow width but similar intensity to normal pulses.

\subsubsection{PSR B2000+40}
PSR B2000+40 was discovered by \citet{Stokes1985} using 92 m Green Bank Telescope at the frequency of 390 MHz. The pulsar has subpulse drifting and periodic nulling, as reported by \citet{Basu2020}. In the FAST GPPS survey, this pulsar is detected by chance. We identify two dwarf pulses from the single-pulse stack of the 5 minute observation on 2021 July 17, and one of them as well as a normal pulse, is displayed in Figure~\ref{FigSinPulStack_J2002_20210717}.

\subsubsection{PSR J2112+4058}
PSR J2112+4058 was discovered by the FAST at 500 MHz \citep{Cruces2021}. This pulsar is detected from the 5 minute data of the FAST GPPS observation on 2021 November 29. From the single-pulse stack, PSR J2112+4058 shows nulling and subpulse drifting behaviors. Three dwarf pulses are detected, and Figure~\ref{FigSinPulStack_J2112_20211129} presents detailed information of a dwarf pulse and a normal pulse.

\section{Discussion and Conclusions}
\label{sect:Discu}

We have detected dwarf pulses in their nulling state for 10 pulsars, in addition to the previously studied cases of PSR J1107-5907 \citep{Young2014} and PSR B2111+46 \citep{Chen2023}. 
A good number of dwarf pulses from PSRs B0525+21, B1237+25, J1851-0053, B1901+10 and B1944+17 have been detected and they are well distinguished from the normal pulses in the distribution of pulse  intensity and pulse width. For the other five pulsars, PSRs J1538+2345, J1824$-$0127, J1939+10, B2000+40 and J2112+4058, a few pulses are detected. In general dwarf pulses have one or more discrete subpulse components. Dwarf pulses usually have a very small energy, often hidden in the distribution peak for nulling periods in the energy distribution histogram of individual pulses. It is possible that very weak emission, even undetectable by FAST, still radiates during the visually nulling periods for some pulsars. Note that the nulling behavior of PSRs J1824-0127, J1939+10 and J2112+4058 is reported here for the first time, with dwarf pulses detected during nulling periods. 

Besides the distinct pulse width and intensity of dwarf pulses from normal pulses, polarization angles of dwarf pulses are often in the orthogonal polarization mode with respect to that of normal pulses. Most dwarf pulses have a very high linear polarization percentage, as seen from the examples presented in plots.
The dwarf emission could appear in any longitude of the on-pulse region defined from the profile of normal pulses, both the core and cone components, as seen for  PSR B1237+25.

As depicted in the diagram of pulsar period ($P$) and period derivative  ($\dot{P}$) in Figure~\ref{FigPPdot}, all these pulsar with such nulling and dwarf pulses are located in the death valley (PSR J1939+10 is not included due to lack of period derivative). This indicates that the nulling and hence the dwarf pulses are probably related to the pair production in a fragile gap of these nearly dead pulsars. 

\begin{acknowledgments}
This work is based on observation data from the FAST. FAST is a Chinese national mega-science facility, built and operated by the National Astronomical Observatories, Chinese Academy of Sciences. %
The authors have been supported by the Natural Science Foundation of China: No. 11988101, 11833009, 12203093 and National SKA Program of China 2020SKA0120100.
\end{acknowledgments}

\bibliography{ref}{}
\bibliographystyle{aasjournal}

\end{document}